\documentclass[11pt]{article}

\usepackage{epsfig}
\def\etal{{\it et al.}}

\begin{document}

{\center{\Large \bf A Cepheid distance to NGC~4258}\\}

\vskip .4in
\normalsize

{\center{Eyal Maoz$^{\#,\ast}$, Jeffrey A. Newman$^{\ast}$, Laura
Ferrarese$^{\dagger}$, Peter B. Stetson$^{\ddagger}$, Stephen
E. Zepf$^{\S}$, Marc Davis$^{\ast}$, Wendy L. Freedman$^{\parallel}$, and
Barry F. Madore$^{\parallel,\P}$}\\}
\vskip 24pt
{\noindent e-mail: {\tt maoz@astro.berkeley.edu, jnewman@astro.berkeley.edu,
lff@astro.caltech.edu, Peter.Stetson@hia.nrc.ca, zepf@astro.yale.edu,
marc@astro.berkeley.edu, wendy@ociw.edu, barry@ipac.caltech.edu}}

\vskip 24pt

\noindent $^{\#}${\it NASA Ames Research Center, MS 245-3, Moffett Field, CA 
94035-1000}\\
$^{\ast}${\it Department of Astronomy, University of California, Berkeley, CA 94720} \\
$^{\dagger}${\it Hubble Fellow, California Institute of Technology,  MS 105-24, Pasadena, CA 91125}\\
$^{\ddagger}${\it Dominion Astrophysical Observatory, 5071 W. Saanich Rd., Victoria, B.C., Canada V8X 4M6}\\
$^{\S}${\it Department of Astronomy, P.O. Box 208101, Yale University, New Haven, CT 06520}\\
$^{\parallel}${\it Observatories of the Carnegie Institution of Washington, 813 Santa Barbara St., Pasadena, CA 91101}\\
$^{\P}${\it NASA/IPAC Extragalactic Database, Infrared Processing and Analysis Center, Jet Propulsion Laboratory, California Institute of Technology, MS 100-22, Pasadena, CA 91125}\\

%\newpage
\vskip .3in

{\bf Distances measured using Cepheid variable stars have been
essential for establishing the cosmological distance scale and the
value of the Hubble constant.  These stars have remained the primary
extragalactic distance indicator since 1929 because of the small
observed scatter in the relationship between their pulsation period
and luminosity, their large numbers, which allow many independent
measures of the distance to a galaxy, and the simplicity of the basic
physics underlying their variability.  Potential systematic
uncertainties in the use of the Cepheid period-luminosity relation to
determine distances are estimated to be $8-10\%$$^{\cite{mould99}}$.
Here we describe the results of a search for Cepheids in the nearby
galaxy NGC~4258, which has an independently determined geometric
distance of 7.2$\pm$0.5~Mpc $^{\cite{herrnstein99}}$.  We determine a
Cepheid distance of 8.1$\pm$0.4 (excluding possible systematic errors
affecting Cepheid distances) Mpc; there is a 1.3$\sigma$ difference
between the measurements. If the maser-based distance is adopted and
other Cepheid distances are revised according to our results, the
derived value of the Hubble constant would be increased by 12$\pm$9\%,
and the corresponding age of the Universe would decrease by the same
factor.}
%\newpage
\vskip .3in

Over the last 20 years, substantial improvements have been made in the
Cepheid distance scale$^{\cite{mf91,jacoby92,wfcrit}}$.  The Cepheid
period-luminosity (P--L) relation derived from observations of stars
in the Large Magellanic Cloud (LMC) has formed the basis for the
calibration of the HST Key Project on the Extragalactic Distance Scale
($^{\cite{keyproject}}$ henceforward, the Key Project) and of many
other recent extragalactic distance measurements$^{\cite{saha94}}$.
Cepheid distances to galaxies as far away as 25 Mpc have been reliably
measured using the Hubble Space Telescope (HST) with random errors
of a few percent; the results have been used to calibrate a number of
secondary distance methods and thereby to provide estimates of the
Hubble constant.

Today, the largest identified potential sources of systematic error in
the Cepheid distance scale are the zero point of the Cepheid
period-luminosity relation (or alternatively, the adopted distance to
the LMC), difficulties in the calibration of the HST Wide Field and
Planetary Camera 2 (WFPC2) camera which has been used for most modern
studies, and the possible effects of differences in the chemical
composition of stars on Cepheid distance measurements.  These have
been judged to be $\pm$ 6.5\%, $\pm$ 4.5\%, and $\pm$ 4\% in distance,
respectively$^{\cite{mould99}}$, with substantial non-Gaussianity
possible; for instance, recently published measurements of the
distance to the LMC range from about 40 to about 55 kpc, with a
distribution skewed towards lower values$^{\cite{wlfrev99}}$.  Given
the remaining uncertainties affecting the application of the
LMC-calibrated Cepheid period-luminosity relation to galaxies observed
with HST, further tests of the extragalactic distance scale are
clearly important.

The spiral galaxy NGC~4258 presents new opportunities to test and
potentially to improve the calibration of the Cepheid P--L relation
because of the precision with which its distance has been measured in
a manner independent of the conventional ladder of astronomical
distance scales$^{\cite{wlfreview}}$.  This distance, $7.2$ Mpc, has
been determined using its apparently simple, Keplerian circumnuclear
disk delineated by line-emitting water masers that orbit a
supermassive black hole at its
center$^{\cite{miyoshi95},\cite{maoz95}}$.  The total estimated
uncertainty in this distance is $\pm 0.3$ Mpc if the disk is presumed
to be circular; if nonzero eccentricities are allowed, the uncertainty
increases to $\pm 0.5$ Mpc (we adopt this value for all further
discussion).  The direct, geometric methods used are believed to have
minimal unknown systematic uncertainties.  Combining the observed
rotation velocities with the measured centripetal acceleration in the
disk$^{\cite{greenhill95}}$ or the observed proper motions of the
maser sources$^{\cite{herrnstein99}}$ allows independent measurements
of the physical size of the disk; comparing these to its observed
angular extent yields the distance to the galaxy center via simple
geometry.  The two routes to this distance (proper motions and
accelerations) yield results in agreement with each other to $1\%$.

We have therefore observed a portion of NGC~4258 to search for
Cepheids using the WFPC2 camera and the Hubble Space Telescope on 11
epochs in 1998 with both the $F555W$ and $F814W$ filters.  Details of
the observations and the analysis procedures summarized here will be
described elsewhere (Newman \etal\ 2000, in preparation); full
photometric results and other additional figures and tables are also
given there.  We have performed point-spread-function fitting
photometry on these data with two commonly-used software packages --
DAOPHOT/ALLFRAME$^{\cite{stetson94}}$ and DoPHOT$^{\cite{saha94,dophot}}$--
following Key Project-like procedures throughout$^{\cite{dophotsearch}}$,
and converted results to standard $V$ and $I$ filter photometric
systems.  Cepheid searches were performed using two different
algorithms with the ALLFRAME dataset$^{\cite{stetson98,newman99}}$,
and using a third for DoPHOT data$^{\cite{dophotsearch}}$.  

We have identified and determined light curves, periods, mean
magnitudes, and colours for 15 Cepheids.  All of these stars fulfill
four criteria: they are identified as variable by all three search
techniques, fit a Cepheid template light curve with reasonable
$\chi^2$, visibly vary in blink comparisons in both $F555W$ and
$F814W$ images, and have negligible statistical probability of being
misidentified nonvariables.  Light curves illustrating the variability
of several of the Cepheids found are plotted in Figure 1.  The DoPHOT
NGC~4258 P--L relations are plotted in Figure 2.  We have adopted the
DoPHOT photometry for all major conclusions reported here, as ALLFRAME
photometry yielded internally discrepant distances from Cepheids on
the two WFPC2 chips used (by 0.24 mag, a 1.8$\sigma$ difference
possibly due to an error in aperture corrections), whereas the
corresponding DoPHOT results agreed to within 0.07 mag (0.5$\sigma$);
however, as an additional check we also provide ALLFRAME values in
much of what follows.

In Key Project procedures, the differences between the observed and
intrinsic colours of Cepheids are then used to correct for the effects
of extinction by dust assuming a standard Galactic reddening
law$^{\cite{cardelli}}$.  Applying this correction star-by-star and
then robustly averaging yields results for small datasets that best
match those that would be obtained using Key Project methods with more
Cepheids (Newman \etal\ 2000, in preparation), so we adopt this as our
primary method of determining a distance modulus for NGC~4258. This
technique yields a distance with a statistical uncertainty (derived
from the P--L relation fit) of $\pm 0.07$ mag.  Based upon past
experience with Key Project galaxies, we would expect a possible
1$\sigma$ systematic error of $\pm 0.04$ mag from the uncertainties in
photometry due to the difficulty of obtaining accurate aperture
corrections in our fields, which contained relatively few bright,
isolated stars.  This error should be highly correlated between $V$
and $I$ measurements and thus not propagated through the reddening
correction.  However, to be conservative we instead adopt the entire
ALLFRAME-DoPHOT distance modulus difference (0.10 mag) as an estimate
of possible errors affecting only photometry for NGC~4258.  Adding
this in quadrature to the random error, the total uncertainty unique
to our determination of a Cepheid distance to NGC~4258 is $\pm 0.12$
mag.

This measurement is subject to a number of potential sources of
systematic error that also affect Key Project distance determinations,
as described in Table 1; their possible contributions have been
estimated to total $\pm 0.18$ mag$^{\cite{ferrarese99}}$. Further
description of these sources of error may be found in Key Project
papers$^{\cite{madore99}}$. Correcting for differences in heavy
element content between the field we have studied and the
LMC$^{\cite{kennicutt98}}$ using previously published studies of
NGC~4258$^{\cite{zkh}}$ would lead to an increase of $0.08 \pm 0.06$
mag in the distance modulus we have derived.  However, due to the lack
of agreement on the magnitude and sign of this effect in recent
studies, the Key Project has refrained from applying such corrections
but considered the resulting uncertainty to be a possible systematic
error.  We thus do likewise, and obtain a Cepheid distance modulus to
NGC~4258 of $29.54 \pm 0.12$ mag (unique to this determination) $\pm
0.18$ mag (systematic uncertainties in Key Project distances),
corresponding to a metric distance of 8.1 $\pm 0.4$ Mpc $\pm 0.7$ Mpc.
When treated in the same way, the ALLFRAME results yield a distance
modulus of $29.64 \pm 0.09$ mag, corresponding to a metric distance of
$8.5 \pm 0.5$ Mpc.  The distance to NGC~4258 derived from observations
of Cepheids is thus appreciably greater than the maser distance of 7.2
$\pm 0.5$ Mpc$^{\cite{herrnstein99}}$.

The Cepheid and maser distances differ by 1.0$\sigma$ if we add our
measurement uncertainty of 0.4 Mpc, the Key Project systematic error
estimate of 0.7 Mpc, and the maser distance error estimate of 0.5 Mpc
in quadrature; potential systematic errors in either technique do not
seem to have been grossly underestimated. If we assume that the maser
distance is correct (up to its stated uncertainty) and wish to
determine whether a revision of the Cepheid distance scale may be
desirable, however, prior estimates of systematic errors in the
Cepheid distance scale are irrelevant to calculations of significance.
In that case, there is a 1.3$\sigma$ (80\% significance) discrepancy
(1.6$\sigma$ if the masing disk is presumed to be circular).  The
largest potential source of such a difference in either error budget
is the uncertainty in the distance modulus to the LMC; however, other
identified potential sources of systematic error are great enough that
a revision of the LMC distance is not required by our results.

A number of other tests of the extragalactic distance scale have been
attempted in recent years using independent, physical or geometric
methods; however, none of them currently have the same precision and
direct applicability to recent Cepheid measurements that the maser
distance to NGC~4258 may provide in combination with our work.
Observations of a ``light echo'' from Supernova 1987A have been used
to place upper limits on the distance to the LMC; however, results
differing by 10\% in distance from each other have been obtained with
this method$^{\cite{gould, panagia}}$, and it cannot place constraints
on errors related to WFPC2 calibration or the effects of differences
in chemical composition from the LMC.  Interferometric radio
observations of the expansion of SN1993J in M81 have been used to
measure its distance with $\sim 15\%$ uncertainty under modest
assumptions $^{\cite{bartel}}$; the agreement with the HST Cepheid
distance to its host galaxy is excellent$^{\cite{kpm81}}$.
However, the uncertainty in the distance to this supernova exceeds the
estimated systematic uncertainties in the Cepheid distance scale, and
the Cepheid distance to M81 was obtained with a different instrument
(WF/PC) and photometric calibration techniques than were used for the
majority of Key Project galaxies.  Finally, distances to several Key
Project galaxies have been obtained using the Expanding Photospheres
Method on supernovae they hosted, again with good agreement.  However,
all of those supernovae were abnormal and considered {\it a priori} to be
poor candidates for the method used; therefore, there may be
substantial systematic errors in those EPM measurements, and drawing strong
conclusions from this agreement may be questionable$^{\cite{epm}}$.

Under the assumption that the maser distance and the estimates of its
uncertainty are correct, we can derive the resulting absolute
magnitudes of the Cepheids observed and obtain a potential
recalibration of the Cepheid P--L relation; the details of such an
analysis will be described in another paper (Newman \etal\ 2000, in
preparation).  Such a calibration could be significantly less subject
to systematic effects than one based on ground-based observations of
the LMC, as Cepheids in NGC~4258 should have a chemical composition
similar to those in galaxies used by the Key Project to calibrate
other distance indicators and have been observed with the same
instruments, filters, and parameter measurement techniques used in
other HST studies.  To lowest order, extinction-corrected Key
Project-like HST WFPC2 Cepheid results would be revised using the
NGC~4258 maser distance simply by subtracting $0.25 \pm 0.19$ mag from
their distance moduli ($0.33 \pm 0.20$ mag if a correction for
differences in heavy element content from the LMC were made to the
NGC~4258 Cepheid distance).  Because the chemical composition of the
field studied in NGC~4258 matches the average for the Key Project
galaxies used to calibrate most secondary indicators, if distances to
both are corrected for differences from the LMC and then revised so
that the Cepheid distance to NGC~4258 would match the maser distance,
the net correction would remain 0.25 mag.  The results for NGC~4258
thus would imply that typical HST WFPC2 Cepheid distances may be too high by
$12\% \pm 9\%$ ($18\% \pm 9 \%$ if the ALLFRAME results were adopted).
It may be worthwhile to note that if applied to all WFPC2 Cepheid
distances, such a correction would eliminate the discrepancies between
the Tully-Fisher relation calibrated using ground-based distances and
that obtained using Key Project results$^{\cite{sakai99}}$, reducing
its overall scatter accordingly.  If such a comparison is fair, this
suggests that the difference in the maser and Cepheid distances may be
more likely to be due to the difficulty of WFPC2 calibration than to
an error in the assumed LMC distance.  However, selection effects or
systematic errors in measuring the total galaxy magnitudes of the
ground-based calibrators, which are relatively large in angular
extent, could also explain the Tully-Fisher discrepancy, as those
magnitudes are required for determining such distances.  Revising
Cepheid-calibrated distance indicators according to the maser distance
would {\bf increase} the measured Hubble constant by $12\% \pm 9\%$,
and {\bf decrease} the corresponding age of the Universe similarly.
The resulting values would be increasingly difficult to reconcile with
globular cluster ages unless the Universe has very low density or,
particularly, if it has a nonnegligible cosmological constant or
similar negative-pressure component$^{\cite{cosmologyimplications}}$.

However, the statistical significance of such a revision remains
limited.  A more compelling test of the HST Cepheid distance scale
based on the maser distance to NGC~4258 would require a substantially
larger sample of Cepheids (reducing the uncertainties in determining the
$VI$ P--L relations and the reddening) and better determination of
aperture corrections; these issues can be addressed simultaneously by
searching for Cepheids with HST in a field that contains more stars
and has undergone more recent star formation.  It would be reasonable
to expect that observations of a region richer in Cepheids might yield
3 times as many Cepheids (giving a distance modulus uncertainty of 0.04
magnitude) and aperture corrections accurate to $\pm 0.04$ magnitude;
better agreement between ALLFRAME and DoPHOT analyses might also
occur with improved data. Reductions of uncertainties in the maser
distance, e.g. via improved constraints on the eccentricity of the
circumnuclear disk, would also be greatly beneficial for its use to
calibrate the extragalactic distance scale.  Successful maser
distances to other galaxies, establishing a Hubble relation, would
more firmly establish this novel technique.  With improvements in both
Cepheid and maser analyses, NGC~4258 has great potential for
establishing a new primary step in the distance ladder, reducing the
potential systematic errors in measurements of the Hubble constant to
perhaps as little as 5\%.

\vskip 0.4in

%\eject 
\pagestyle{empty}

\vskip 0.3in
%\eject
\begin{center}
{\center \bf Figure and Table Captions}
\end{center}
\vskip 12pt
\noindent {\bf Figure 1} -- Light curves for three representative
examples of the Cepheids we have discovered in NGC~4258.  Open symbols
depict the measured magnitudes of a given star, and the error bars the
uncertainty therein; the solid lines show template Cepheid light
curves derived from observations of stars in the
LMC$^{\cite{stetson96}}$ that have been used to fit the data.  The
left panel for each star shows the variation of the $V$ magnitude
of each star versus the the phase of the variation, which makes the
Cepheid pulsation pattern clear.  The right panel
plots the $I$ variation of the Cepheids in similar fashion.
\vskip 24pt

\noindent{\bf Figure 2} -- NGC~4258 Period--Luminosity relations.  {\bf
a}, The $V$ P--L relation for the Cepheids we have found in NGC~4258,
based upon DoPHOT photometry.  The Cepheid P--L relation has
previously been found to be well fit by a linear relationship between
mean magnitude and the logarithm of the period in the range 10--60
days.  To minimize the effects of incompleteness and to optimize
results given the limited number of Cepheids observed, we adopt LMC
P--L relation slopes$^{\cite{mf91}}$ and only fit for differences in
the zero point.  Open symbols depict the measured parameters of
NGC~4258 Cepheids; the solid line indicates the fit P--L relation, and
the dashed lines the 2$\sigma$ observed scatter of fiducial LMC
Cepheids about its P--L relation.  From the difference between the
absolute magnitudes of LMC Cepheids (for an assumed LMC distance of 50
kpc) and the observed magnitudes of NGC~4258 Cepheids we derive a $V$
distance modulus, defined by $\mu = 5 {\rm log_{10}} d / 10 {\rm pc}$,
of $\mu_V$=$29.82 \pm 0.07$ (ALLFRAME results yield $29.88 \pm 0.08$).
This value must be corrected for the effects of dust extinction to
yield a reliable measurement of the distance of NGC~4258.  {\bf b},
The $I$ P--L relation for NGC~4258 Cepheids.  Applying procedures
similar to those for $V$, we derive an $I$ distance modulus
$\mu_I$=$29.71 \pm 0.05$ (ALLFRAME photometry yields $29.78 \pm
0.05$).  Correcting for the effects of dust on the apparent magnitude
of each star using its observed colour, we obtain an
extinction-corrected distance modulus of $29.54 \pm 0.07$ mag from
DoPHOT photometry for all Cepheids; those on WFPC2 chip 2 alone yield
$29.58 \pm 0.12$, while those on chip 3 yield $29.51 \pm 0.09$ (for
ALLFRAME photometry, the corresponding numbers are $29.64 \pm 0.08$,
$29.50 \pm 0.12$, and $29.74 \pm 0.10$ mag).
\vskip 24pt

\noindent{\bf Table 1} -- A summary of the major identified errors which
may affect our extinction-corrected distance modulus measurement.  All
numbers listed are in magnitudes; for small differences, the
corresponding fractional error in distance may be obtained by
multiplying by 0.46.  For those potential systematic errors which
affect all Cepheid distances obtained in the same manner as ours
uniformly, we have adopted the uncertainty estimates of the Key
Project$^{\cite{madore99}}$; more detailed descriptions may be found
therein.  The estimated systematic errors in WFPC2 photometry (S$_2$)
include uncertainties in zero points and the `long versus short'
uncertainty, added in quadrature.  Because the metallicity of the
field we have studied in NGC~4258 matches the average metallicity of
Key Project galaxies used to calibrate the Tully-Fisher relation,
surface brightness fluctuation distances, and the peak luminosities of
Type Ia supernovae, we treat the uncertainties in metallicity
correction as a systematic error common to this distance measurement
and those of the Key Project (S$_3$).  The random uncertainty in the
Cepheid distance modulus (R$_1$) has been estimated from the standard
error of the extinction-corrected moduli for individual stars.  We
have adopted a conservative estimate of the total photometric
uncertainties unique to NGC~4258 (S$_4$) of 0.10 mag based on the
difference between the overall distance moduli they yield.  For chip
2, the mean difference between ALLFRAME and DoPHOT magnitudes for 24
bright, isolated stars was $0.026 \pm 0.049$ (standard deviation) mag
for $V$, and $0.015 \pm 0.049$ mag for $I$.  For chip 3, the mean
difference for 30 stars was $0.025 \pm 0.027$ mag for $V$, and $0.088
\pm 0.046$ mag for $I$.  Mean magnitudes for Cepheids yielded results
consistent with these to within 1$\sigma$, albeit with much larger
standard errors.  We consider the chip 3/$I$ results to be an
aberration closely related to the discrepant distance moduli obtained
with ALLFRAME on the two chips; even including it, the mean
ALLFRAME-DoPHOT magnitude difference among the 4 cases would be 0.04
mag.

\newpage
\begin{figure}[t]
\vskip -2in
\epsfbox{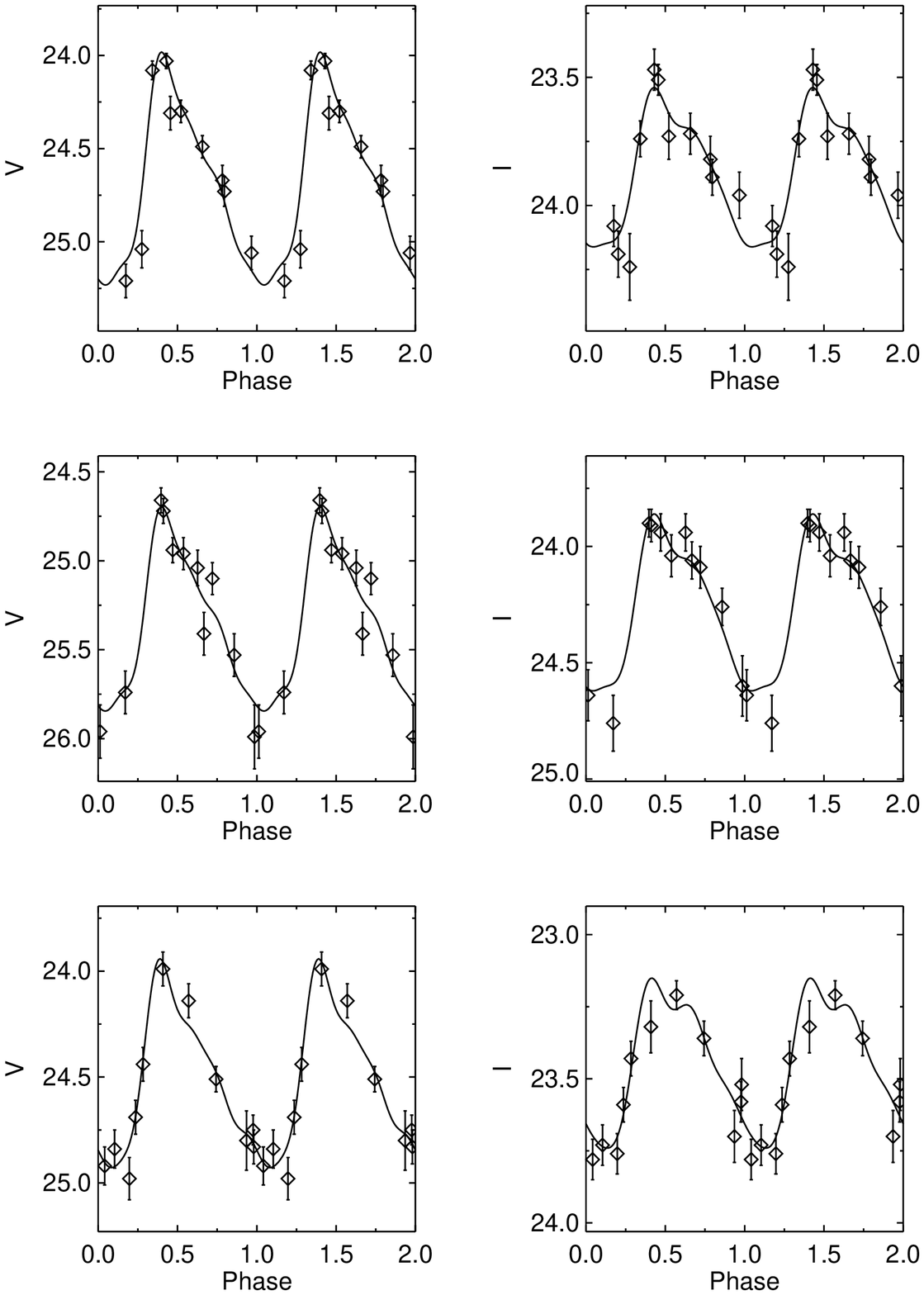}
\vfill
\begin{center}
{\bf\center Figure 1}
\end{center}
\end{figure}

\newpage
\begin{figure}[t]
\vskip -2in
\hskip -.5in
\epsfbox{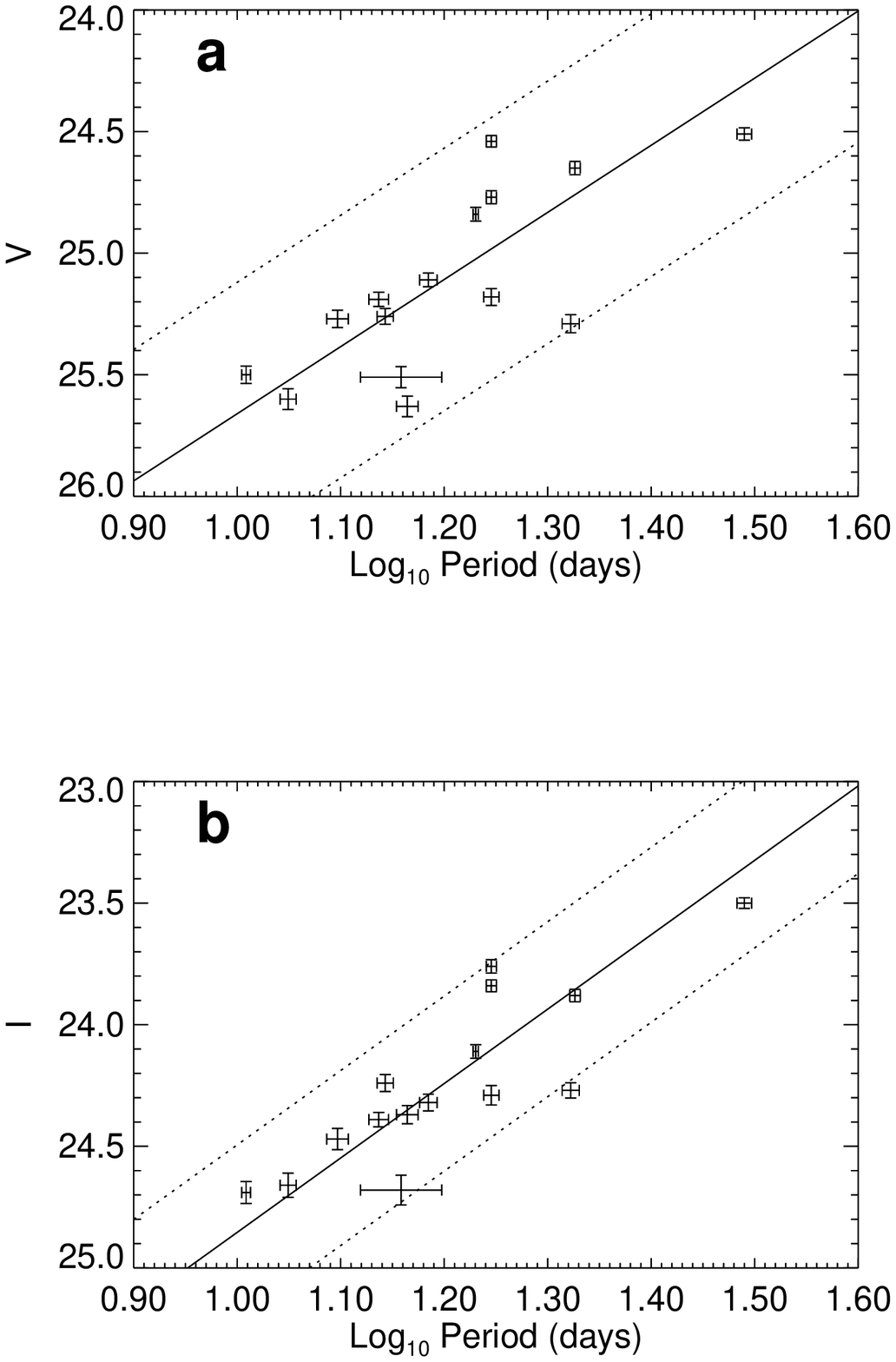}
\vfill
\begin{center}
{\bf\center Figure 2}
\end{center}
\end{figure}

\newpage
\begin{figure}[t]
\vskip -2in
\hskip -.9in
\epsfbox{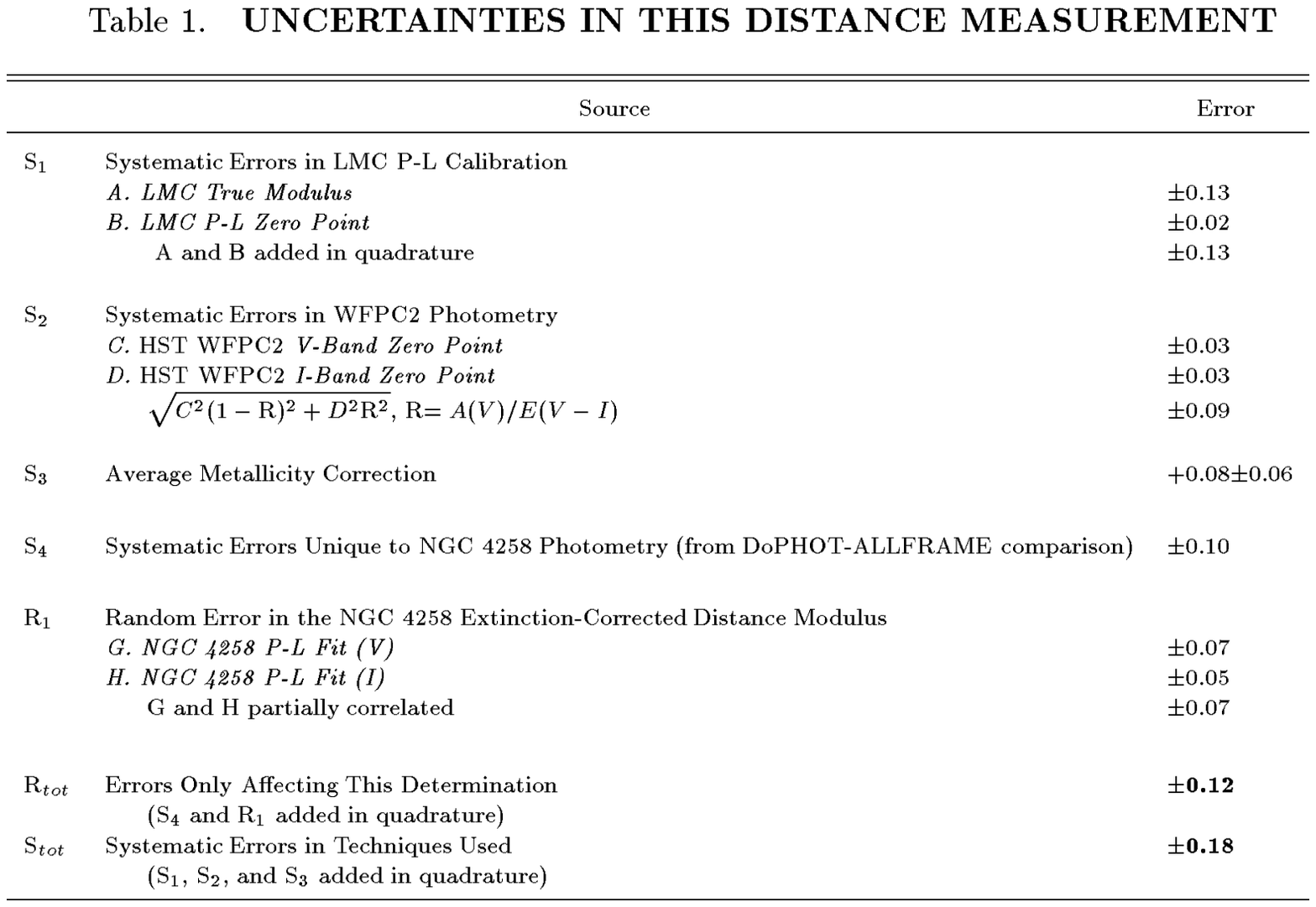}
\vfill
\end{figure}

\end{document}